\documentclass[aps,pra,superscriptaddress,12pt,tightenlines,nofootinbib]{revtex4}
\usepackage{amsmath,amsthm,graphicx,amssymb,verbatim,listings}
\usepackage[T1]{fontenc}     
\usepackage{lmodern}         
\usepackage[ pdftex, plainpages = false, pdfpagelabels,
                 pdfpagelayout = useoutlines,
                 bookmarks,
                 bookmarksopen = true,
                 bookmarksnumbered = true,
                 breaklinks = true,
                 linktocpage=all,
                 pagebackref=false,
                 colorlinks = true,
                 linkcolor = BrickRed,
                 urlcolor  = blue,
                 citecolor = BrickRed,
                 anchorcolor = green,
                 hyperindex = true,
                 hyperfigures
                 ]{hyperref}
\usepackage[usenames, dvipsnames]{xcolor}
\usepackage{xifthen} 

\newcommand{\arxiv}[2][]{\ifthenelse{\isempty{#1}}{\href{http://arxiv.org/abs/#2}{{\tt arXiv:\allowbreak{}#2}}} {\href{http://arxiv.org/abs/#2}{{\tt arXiv:\allowbreak{}#2 [#1]}}}}

\newcommand{\booktitle}{\textsl}
\newcommand{\hrefdoi}[2]{\href{https://dx.doi.org/#1}{#2}}

\newcommand{\defterm}[1]{\textbf{#1}}

\begin{document}

\title{Multiscale Structure of More-than-Binary Variables}

\author{Blake C.\ Stacey}
\affiliation{Physics Department, University of Massachusetts Boston}

\date{\today}

\begin{abstract}
In earlier work, my colleagues and I developed a formalism for using
information theory to understand scales of organization and structure
in multi-component systems.  One prominent theme of that work was that
the structure of a system cannot always be decomposed into pairwise
relationships.  In this brief communication, I refine that formalism
to address recent examples which bring out that theme in a novel and
subtle way.  After summarizing key points of earlier papers, I
introduce the crucial new concept of an \emph{ancilla component,} and
I apply this refinement of our formalism to illustrative examples.
The goals of this brief communication are, first, to show how a simple
scheme for constructing ancillae can be useful in bringing out
subtleties of structure, and second, to compare this scheme with
another recent proposal in the same genre.
\end{abstract}

\maketitle

Imagine a system that is made up of a large number of pieces.  A theme
one encounters in many areas of science is that such a system is
simpler to understand when the pieces are independent of one another,
each one ``doing its own thing.''  In thermodynamics and statistical
physics, for example, the study of fluids begins with the conceptual
model known as the ideal gas, in which the atoms sail past each other
without interacting.  On the other hand, a system can also be simple
to understand if all the pieces are so tightly correlated or so
strongly bound together that they essentially move as a unit.  Then,
the task of understanding the system again simplifies, because knowing
what any one part is doing tells us most of what we need to know about
the whole.

The general approach that my colleagues and I have developed over
recent publications is that we can use \emph{information theory} to
make these heuristic discussions quantitative~\cite{allen2014,
  stacey-thesis, complexityinf}.  One conclusion is that we should not
aim to quantify a system's ``complicatedness'' by a single number.
Instead, it is more illuminating to devise \emph{curves} that indicate
how much structure is present \emph{at different scales.}  Another
general theme is that one can mathematically formalize the notion of
describing a system by augmenting that system with extra components
and applying information theory to the augmented system.  In this
note, I will review the relevant aspects of information theory and the
formal concept of system description, and then I will refine our
earlier treatment of system description~\cite{allen2014} to make it
sensitive to complex structure in new ways.

We start by saying that a system $\mathcal{S}$ is composed of pieces,
or components.  We will denote the set of components of the system
$\mathcal{S}$ by $S$.  It is important to distinguish the two, because
we could take the same components and arrange them in a different way
to create a different system.  We express the arrangement or the
patterning of the components by defining an \defterm{information
  function}. For each subset $T \subset S$, the information function
$H$ assigns a nonnegative number $H(T)$, which expresses how much
information is necessary to describe exactly the configuration or the
behavior of all the components in the set $T$.  One can prove a
significant amount from the starting point that if $H$ is to be an
information function, it must satisfy a few basic axioms.

First, as we stated above, 
\begin{equation}
H(T) \geq 0 \hbox{ for any } T \subset S.
\end{equation}
Second,
\begin{equation}
\hbox{if } T \subset V \subset S,\hbox{ then } H(T) \leq H(V).
\end{equation}
We call this property \defterm{monotonicity}.

Third, if we have two subsets $T \subset S$ and $V \subset S$, the
total information assigned to their union, $H(T \cup V)$, must be
limited.  Information pertinent to the components in $T$ can be
pertinent to the components in $V$.  For one reason, the sets $T$ and
$V$ might have some components in common.  And even those components
that are not shared between the two sets might be correlated in some
way such that the amount of information necessary to describe the
whole collection is reduced.  So, we require that
\begin{equation}
H(T \cup V) \leq H(T) + H(V) - H(T \cap V),
\end{equation}
which we call \defterm{strong subadditivity}.  Given two components $s$
and $t$ within $S$, the \defterm{shared information} expresses the
difference between what is required to describe them separately versus
describing them together:
\begin{equation}
I(s;t) = H(\{s\}) + H(\{t\}) - H(\{s,t\}).
\label{eq:shared-info}
\end{equation} 
It follows from strong subadditivity that the shared information is
always nonnegative.

In the full multiscale formalism~\cite{allen2014}, each component $s
\in S$ also has associated with it a \defterm{scale}, $\sigma(s)$,
which is a quantitative measure of that component's significance or
importance.  This allows us to discuss the scale at which information
applies.  For the examples we will consider in this note, each
component will be of equal intrinsic significance, so $\sigma(s)$ will
be normalized to unity for all $s$.

A \defterm{descriptor} of a system is an entity outside that system
which tells us about it in some way.  Mathematically speaking, we take
the system $\mathcal{S}$ and augment it with a new component that we
can call $d$, to form a new system whose component set is $S \cup
\{d\}$.  The information function of the augmented system, which we
can denote $H^\dag$, reduces to that of the original system
$\mathcal{S}$ when applied to subsets of $S$.  That is, $H^\dag(T) =
H(T)$ for all $T \subset S$.  The original information function $H$ is
not defined on sets that include the descriptor $d$, but the
information function $H^\dag$ of the augmented system is.  The values
of $H^\dag$ on subsets of $S \cup \{d\}$ that include the descriptor
$d$ tell us how $d$ shares information with the original components of
$\mathcal{S}$.  Just like $H$, the function $H^\dag$ obeys the axioms
of monotonicity and strong subadditivity, and we can construct a
shared information $I^\dag(s;t)$ following Eq.~(\ref{eq:shared-info}).

The \defterm{utility} of a descriptor $d$ is measured in information
weighted by scale:
\begin{equation}
U(d) = \sum_{s \in S} \sigma(s) I^\dag(d;s).
\end{equation}
Given the basic axioms of information functions that we listed above,
we can define an \defterm{optimal descriptor} as the one which has the
largest possible utility, given its own amount of information.  That
is, if we invest an amount of information $y$ in describing the system
$\mathcal{S}$, then an optimal descriptor has $H^\dag(d) = y$, and it
relates to $\mathcal{S}$ in such a way that $U(d)$ is as large as the
basic axioms of information functions allow.  This defines a linear
programming problem whose solution is the \defterm{optimal utility},
and so the theory of linear programming lets us prove helpful results
about how the optimal utility varies as a function of $y$.  Taking the
derivative of the optimal utility gives the \defterm{marginal utility
  of information}, or MUI.

We proved several useful properties of the MUI in an earlier
article~\cite{allen2014}.  For systems of the ideal-gas type, where
information applies to one component at a time, the MUI is a low and
flat function: Investing one bit of description buys one unit of
utility, until the whole system is described.  On the other hand, if
all the components are bound together and ``move as one,'' then
investing a small amount of information buys us utility on a large
scale, because that small amount applies across the board.  In this
case, the MUI starts high and falls sharply.

The MUI is defined using a single descriptor component.  It is natural
to speculate that constructions involving multiple descriptors could
provide useful elaborations of the multiscale complexity formalism.
This is one motivation for the developments that follow.

When the construction of a system is specified in detail, it is
sometimes possible to make a finer degree of analysis, which reveals
features that a first application of a structure index can overlook.
To illustrate this, consider a system defined by a set
of random variables, to which we ascribe some joint probability
distribution.  When systems are defined in this way, we can use the
\defterm{Shannon information} (a.k.a., Shannon entropy, Shannon index)
as our information function $H$.  In the absence of an external
reference to compare the values of these variables against, the most
obvious meaningful statement we can make about them is whether the
values are equal.  (If the numbers recorded masses in grams, for
example, then the difference between ``0'' and ``3'' would be more
dramatic than that between ``0'' and ``1,'' but we do not know that
\emph{a priori.}  Information theory has, in certain ways, neglected
the idea that some differences between symbols are more striking than
others~\cite{allen2009, leinster2012}.)  Let us say that the system
has three components, $a_1$, $a_2$ and $a_3$.  Following the general
idea of adding a descriptor to the system, as we did with the MUI, we
introduce a new variable $\Delta_{12}$ which takes the value 1 when
the state of $a_1$ and $a_2$ are the same, and is 0 otherwise.  This
new \defterm{ancilla} variable is determined completely by the
original system, and is sensitive to the particular values taken by
the original system components $a_1$, $a_2$ and $a_3$.  We can define
two other ancillae in the same way, $\Delta_{13}$ and $\Delta_{23}$.
Then, we can use the tools of multiscale information theory, like the
MUI, to explore the structure of the ancilla variables, which in turn
tells us about the structure of the original system.  (The term
``ancilla'' is common in \emph{quantum} information theory, in a sense
similar to this~\cite{peres1990, caves2002, nielsen2011}.)

As a preliminary, let us try this with a three-component
\defterm{parity-bit system}, which we can think of as picking a row at
random from the \textsc{xor} truth table:
\begin{equation}
X = \left(\begin{array}{ccc}
0 & 0 & 0 \\
0 & 1 & 1 \\
1 & 0 & 1 \\
1 & 1 & 0
\end{array}\right).
\end{equation}
Then, the values of the ancillary variables $\Delta_{12}$,
$\Delta_{13}$ and $\Delta_{23}$ in each possible joint state are the
rows in the matrix
\begin{equation}
\Delta_X = \left(\begin{array}{ccc}
1 & 1 & 1 \\
0 & 0 & 1 \\
0 & 1 & 0 \\
1 & 0 & 0
\end{array}\right).
\end{equation}
This is, again, a parity-bit system, but with odd parity instead of
even.  We can also think of it as the truth table of the \textsc{xnor}
logic gate: Any column is the \textsc{not} of the \textsc{xor} of the
other two.  The fact that the structure does not simplify when we
consider the variables pair-by-pair indicates that this system is not
structured in a pairwise way, confirming what we noted
in~\cite{allen2014}.

For a more elaborate example, take the two systems defined by James
and Crutchfield~\cite{james2016}.  These are three-component systems
composed of random variables whose joint states are chosen by picking
a row at random from a table, with uniform probability.  The
\defterm{dyadic} and \defterm{triadic} systems are defined
respectively by the tables
\begin{equation}
D = \left(\begin{array}{ccc}
0 & 0 & 0 \\
0 & 2 & 1 \\
1 & 0 & 2 \\
1 & 2 & 3 \\
2 & 1 & 0 \\
2 & 3 & 1 \\
3 & 1 & 2 \\
3 & 3 & 3
\end{array}\right); 
\ T = 
\left(\begin{array}{ccc}
0 & 0 & 0 \\
1 & 1 & 1 \\
0 & 2 & 2 \\
1 & 3 & 3 \\
2 & 0 & 2 \\
3 & 1 & 3 \\
2 & 2 & 0 \\
3 & 3 & 1
\end{array}\right).
\end{equation}
If we compute the MUI for these two systems in the manner described
above, we find that the MUI for the dyadic system is the same as for
the triadic.  In fact, the information functions for the two systems,
computed according to the Shannon scheme, agree for all subsets $U$.
However, we can detect a difference between their structures by
\emph{augmenting} them, in the manner described above.

Specifically, if we take the dyadic example system and introduce three
ancillae $\Delta_{12}$, $\Delta_{13}$ and $\Delta_{23}$ in the manner
described above, we find that the three ancillae form a completely
correlated block system (that is biased towards the joint state 000).
In contrast, for the triadic example, defining three ancillae in the
same way, we find that they form a parity-bit system (with odd
parity).  This reveals that the tradic system has an
information-theoretic structure at the scale of three variables in a
way that the dyadic system does not.  Explicitly,
\begin{equation}
\Delta_D = \left(\begin{array}{ccc}
1 & 1 & 1 \\
0 & 0 & 0 \\
0 & 0 & 0 \\
0 & 0 & 0 \\
0 & 0 & 0 \\
0 & 0 & 0 \\
0 & 0 & 0 \\
1 & 1 & 1
\end{array}\right);
\ \Delta_T = 
\left(\begin{array}{ccc}
1 & 1 & 1 \\
1 & 1 & 1 \\
0 & 0 & 1 \\
0 & 0 & 1 \\
0 & 1 & 0 \\
0 & 1 & 0 \\
1 & 0 & 0 \\
1 & 0 & 0
\end{array}\right).
\end{equation}
The ancillary system defined by $\Delta_T$ has four possible distinct joint
states, all of which are equally probable, and so it has two bits of
information overall.  For each possible joint state of the ancillary
system, the original system can be in one of two joint states, with
equal probability.  Therefore, we see that the three bits of
information necessary to specify the state of the original system
break down into a pair of bits for describing the ancillae, plus one
more bit of additional detail.  

For both the dyadic and triadic examples, the MUI of the ancillary
systems $\Delta_D$ and $\Delta_T$ takes a simple form; in fact, these
cases were both solved in~\cite{allen2014}.  For the ancillae of the
dyadic system, $\Delta_D$, if we define
\begin{equation}
h = -\frac{1}{4}\log_2\left(\frac{1}{4}\right)
 - \frac{3}{4}\log_2\left(\frac{3}{4}\right) \approx 0.8113,
\end{equation}
we have
\begin{equation}
M(y) = \left\{\begin{array}{cc}
3, & y \leq h; \\
0, & y > h.
\end{array}\right.
\end{equation}
And for the ancillae $\Delta_T$ of the triadic system, we have
\begin{equation}
M(y) = \left\{\begin{array}{cc}
\frac{3}{2}, & y \leq 2; \\
0, & y > 2.
\end{array}\right.
\end{equation}

MUI is in general a piecewise linear function.  These two examples are
both rather simple, in that there is only one step; for natural
examples of MUI curves having multiple successive drop-offs,
see~\cite{allen2014}.  The vertical axis of any MUI curve has units of
scale, while the horizontal axis indicates an amount of information,
which in the Shannon case is measured in bits (or alternatively in
nats, etc.).  So, the drops in the MUI curve do not have to occur at
integer values.  This is illustrated nicely by the dyadic example:
Investing 0.8133 bits of information means that you can buy a complete
description of the system, so the marginal return on information
invested drops to zero at 0.8133 bits, because after that point,
there's nothing more to buy.  The numerical value of~$M(y)$, in the
region where $M(y) > 0$, indicates the scale at which the information
in an optimal description applies; for a more in-depth discussion,
see~\cite{allen2014}.

The appeal of the pairwise-comparison ancilla scheme is threefold.
First, it is readily motivated: In the absence of some kind of metric,
the natural and easy thing to do when comparing outcomes is to test
whether they are equal.  Second, it is straightforward to define for
systems of any size.  Third, when applied to the James--Crutchfield
examples, the XOR structure hiding in the ternary example pops out
immediately.  Computing the MUI was, in that case, really an
afterthought: The XOR structure simply appears in the table of the
ancillae's joint states.

The ancilla scheme has a certain conceptual commonality with the
\defterm{connected information profile}, which quantifies the
structure of a joint probability distribution by constructing a
hierarchy of new distributions whose marginals agree with those of the
original~\cite{Schneidman2003}.  When applied to the triadic example,
the connected information profile also identifies two bits of
information in pairwise correlations and one bit of additional
detail~\cite{james2016}.  The relation between these ways of
quantifying system structure deserves, I think, further work.

If $X$ and $Y$ are two independent random variables with the same
probability distribution, then the Shannon information of the pairwise
ancilla $\Delta_{XY}$ is fixed by the probability that an observation
of $X$ and an observation of $Y$ yield the same value.  This
describes, for example, an experiment where we draw a ball from a
well-mixed urn, note its color, return the ball to the urn, stir well
and draw a ball at random again.  In this context, the probability of
observing the same color twice is known as an \defterm{index of
  coincidence}~\cite{qplex}.  Such indices are useful in ecology, for
quantifying population diversity~\cite{leinster2012}; they can also be
interpreted as expected scores in symmetric games whose goal is
agreement~\cite[\S 10.5]{stacey-thesis}.

Since I first posted a note on this topic~\cite{stacey-blog}, Ince
published an article on the arXiv that also addressed the scenarios
posed by James and Crutchfield~\cite{Ince2017}.  Ince's article is a
good occasion to compare our approach to multiscale information (and
the elaboration of it we have begun here) to other programs of
research.  In what follows, I will attempt to draw attention to
certain distinctions, which I believe are better thought of as
complementary, rather than contradictory, conceptual developments.

Another way we could have elaborated upon our earlier multiscale
information formalism would be to define an ancilla for each possible
value of each component of a system.  We could designate this
\defterm{exploding} the original system.  Applying our theory to an
exploded system can reveal new details of organization, at the price
of increasing the number of components one must consider.  Suppose
again that we have $N$ random variables, each one of which has $K$
internal states.  If we explode the system and invent an indicator
variable for each possible internal state of each original variable,
then we have $NK$ indicator variables, and specifying an information
function on that set requires fixing $2^{NK} - 1$ numbers.

For example, consider a two-component system defined by picking two
successive characters at random out of a large corpus of English text.
The shared information between the two components quantifies how much
knowing the value of the first character helps us predict the value of
the second.  However, knowing that the first character is a $Q$ is a
stronger constraint than knowing that it is, say, a $T$, because fewer
characters can follow a $Q$.  We can express this in our theory by
exploding the two-component system, defining new components that
represent the events of the first character being a $Q$ and the first
character being a $T$.  This idea also finds expression in the
\defterm{partial information decomposition}~\cite{wb-pid}.  When
defining the PID, one uses a set of values of ``specific
information,'' which is a measure of the information that a source
variable provides about a particular outcome of a target variable.
The specific information between a source $X$ and a specific value $y$
of a target $Y$, denoted $I(Y=y;X)$, measures the amount by which
learning the value of~$X$ makes observing the value $y$ become less
surprising.  By construction, averaging $I(Y=y;X)$ over all values of
the variable $Y$, weighted by their probabilities, yields the standard
mutual information $I(Y;X)$.  The PID requires the user to introduce a
distinction between ``input'' and ``output'' variables.  Of course, in
some scenarios, such a distinction is a natural one to introduce, but
(as James and Crutchfield note) that is not true everywhere.

The role of the ``specific information'' values in the original PID is
as raw material to construct a \defterm{redundancy} function on sets
of variables.  However, in the years that followed the proposal of the
original PID, it became increasingly clear that its way of quantifying
the redundancy of information was not adequate, and the proper way to
define a redundancy measure is not obvious~\cite{Bertschinger2014,
  Ince2016}.  The role of the scale function in the multiscale
approach---recall that $\sigma(s)$ is provided independently of
$H(U)$---suggests that in practice, it may not always be possible to
define a meaningful measure of redundancy among system components only
using joint probability distributions.

Ince proposes a \defterm{partial entropy decomposition}, inspired by
the PID but without the distinction between ``input'' and ``output''
variables.  Like our treatment with ancilla variables, Ince's PED can
identify the parity-bit structure within the triadic example system.
A key conceptual issue with the PED is that one step in the definition
explicitly requires the use of probabilities, and so it can only be
formulated for the Shannon index.  The same holds true for the
connected information profile mentioned earlier.  As we have argued
elsewhere~\cite{allen2014}, the concept of ``information'' ought to be
considered more generally.  Rather than founding everything on
logarithms of probabilities, it is beneficial to prove as much as
possible starting from \emph{the properties that a reasonable measure
  of information ought to satisfy.}  (This approach is perhaps closest
in spirit to that of Quax, Har-Shemesh and Sloot~\cite{Quax2017},
which likewise appeared in a journal after the first version of this
note was posted~\cite{stacey-blog}.)  This methodology clarifies what
essential features a structure index depends upon, facilitates the
exchange of concepts among areas of mathematics, and tells us what
must be modified when we carry the idea of ``information'' into new
contexts.  (For example, in \emph{quantum} information theory, the way
that probability distributions interlock is
deformed~\cite{qplex, stacey-qutrit, debrota2017}, and while the
natural information function still obeys strong subadditivity, the
monotonicity property must be weakened~\cite{linden2005, gross2013,
  linden2013}.)  It would be interesting to see how the PED might be
formulated in this manner.

Finally, note that Ince develops the PED for three-component
systems, and demonstrates its usefulness there.  As Ince points out,
extensions to larger systems require making some decisions on a
conceptual level, but they should be quite interesting as well.  I
suspect that a PED that is viable for larger systems could open the
way to merging the concepts of the PID/PED and the MUI.  For example,
if one interprets the descriptor component used in defining the MUI as
an ``input'' variable, then the input-output thinking behind the PID
starts to become applicable, suggesting that the PID and PED could
lead to modified definitions of utility that, in the right scenarios,
are fruitful.

\bigskip

My colleagues have been advocating the study of multiscale structure
for a long time~\cite{BarYam1997, BarYam2004, SGSYBY2004}, and so I am
grateful to all those who are raising new challenges in the subject,
thereby demonstrating its intellectual vitality.  In particular, I
thank Robin Ince for email conversations that led to a v2 for this
essay.

\end{document}